\documentclass[aps, pra, twocolumn]{revtex4-2}

\usepackage{amsmath}
\usepackage{amssymb}
\usepackage{xcolor}
\usepackage{url}
\usepackage{siunitx}
\usepackage{graphicx}

\usepackage[english]{babel}
\babeladjust{
    autoload.bcp47 = on
}
\newcommand{\wavenumber}{\per\centi\meter\tothe{1}}

\newcommand{\Fseven}{\ensuremath{{}^2\mathrm{F}_{7/2}}}
\newcommand{\Ffive}{\ensuremath{{}^2\mathrm{F}_{5/2}}}

\begin{document}

\title{Site-Resolved Optical Spectroscopy of Thulium in Solid Argon}

\author{Jason Marfey}
\email{jmarfey3@gatech.edu}
\author{Heming Gao}
\email{hgao96@gatech.edu}
\author{Colin V. Parker}
\email{cparker@gatech.edu}
\affiliation{School of Physics, Georgia Institute of Technology, Atlanta, GA, 30332, USA}

\begin{abstract}
    We utilize the high resolution of the ground state transition ${}^2\mathrm{F}_{7/2} \leftrightarrow {}^2\mathrm{F}_{5/2}$ in thulium atoms isolated in solid argon to perform excitation-emission spectroscopy.
    Decomposition of the excitation-emission spectrum reveals 6 absorption-emission profiles, which are attributed to multiple trapping sites in the host matrix.
    Comparison with vacuum transition energies shows large matrix-induced shifts for $\mathrm{s}\rightarrow\mathrm{p}$ transitions, whereas $\mathrm{f} \rightarrow \mathrm{d}$ transitions remain close to their free-atom energies.
    A single site exhibiting strong emission at \qty{560}{\nano\meter} is investigated and a three-photon excitation pathway is determined as a possible origin of the high energy \qty{560}{\nano\meter} light.
    Additionally, for long-lived levels belonging to the $\mathrm{f}^{12}\mathrm{d}$ configuration, narrow, crystal-field split absorption lines are identified.
    Matrix effects are important for establishing the mechanism behind the observed spectra and long $\mathrm{f}^{12}\mathrm{d}$ lifetimes.
\end{abstract}

\maketitle
\section{Introduction}

Precision measurements of the optical spectrum of an atom provide a fingerprint of its internal structure and a record of its surroundings.
In a rare-gas solid, this sensitivity can be used to probe local fields while the atom remains trapped in a host that can be grown near the surface of interest~\cite{gaire2019,marfey2025}.
Recent state-of-the-art optical system designs have shown that the internal state of a matrix-isolated atom can be accessed even at the single-atom level~\cite{lancaster2024,chambers2019}.
The development of these delicate systems depends on a substantial understanding of how the host changes the atomic spectrum and what the resulting fluorescence tells us about the environment.

In many lanthanide atoms, filled outer shells shield the partially filled $4\mathrm{f}$ shell.
Transitions within this shell can remain narrow or weakly shifted in solid rare gases~\cite{gaire2023,marfey2025,zaitsev2026}.
The narrow lines permit selective excitation, and higher excited states can provide faster fluorescence for readout.
The choice of a fluorescence channel depends on how the host affects the spectra and relaxation of the excited configurations~\cite{lambo2012,lambo2021}.

Thulium is a dopant of interest because its 1140~nm magnetic-dipole transition connects the ${}^{2}\mathrm{F}_{7/2}$ and ${}^{2}\mathrm{F}_{5/2}$ levels of the $4\mathrm{f}^{13}6\mathrm{s}^{2}$ ground configuration, and the shielding of the $4\mathrm{f}$ electrons by closed outer shells reduces the transition's sensitivity to environmental electric fields~\cite{ishikawa1997,golovizin2019}.
Early measurements in solid argon and neon found narrow infrared emission, and subsequent laser spectroscopy resolved multiple components with ensemble linewidths below 1~GHz~\cite{gaire2019,gaire2023}.
These narrow resonances provide a basis for selective optical excitation and upconversion pumping of thulium.
Previous pump-probe spectroscopy of thulium identified two thermally stable trapping sites in solid argon, labeled Site I and Site II, and their crystal-field level structure was determined~\cite{marfey2025}.
Several visible bands were observed from Site I, whereas no visible emission was detected from Site II under the conditions of those measurements.

In our experiment, pulsed laser ablation supplies Tm from a solid target as Ar condenses on a cold substrate, allowing the dopant supply to be controlled separately from the Ar flow~\cite{marfey2025}.
The trapping environments observed in a spectrum can depend on sample preparation, as seen previously in laser-ablated Li in solid Ar and Kr, which exhibits trapping-site features absent from Knudsen-effusion controls~\cite{fajardo1991}.
Classifying the possible trapping sites for a dopant is important for quantitative studies, particularly for experiments aiming for precision physics tests~\cite{battard2023,lahs2026,vutha_orientation-dependent_2018}.
In Tm samples prepared by laser ablation, excitation resonances from different sites can overlap near 1140~nm, and emission bands from different sites can fall in the same spectral region.
Site-selective excitation spectroscopy of Eu in Ar and Kr separated overlapping site contributions and enabled a reassessment of the crystal-field structure of $\mathrm{f}\rightarrow\mathrm{d}$ bands in Ar~\cite{byrne2011}.
Selective excitation also helped separate visible emission from the Eu sites in Ar and Kr~\cite{byrne2011luminescence}.
We measure the Tm fluorescence as a function of both excitation and emission wavelength.
In the resulting excitation--emission map, differences in excitation dependence help distinguish nearby bands, and bands at different wavelengths can be grouped by a shared excitation profile.

With the emission bands grouped by site, comparison with free-atom transition energies helps assign the higher-lying electronic states.
These assignments depend on how the matrix affects the higher-lying configurations involved in visible emission.
Optical absorption measurements of neutral Sm in solid Ar found band splitting and blue shifts of up to $800~\mathrm{cm}^{-1}$ relative to the gas phase~\cite{klotzbucher1994}.
Yb in solid Ne exhibits broad $6\mathrm{s}\rightarrow6\mathrm{p}$ absorption bands and much narrower $4\mathrm{f}\rightarrow5\mathrm{d}$ bands with resolved crystal-field and phonon structure~\cite{lambo2021}.
Recent measurements of neutral Er in solid Ne and Ar likewise show weakly shifted $4\mathrm{f}\rightarrow4\mathrm{f}$ emission alongside more strongly structured emission involving $5\mathrm{d}$ states~\cite{zaitsev2026}.
These observations motivate examining the $4\mathrm{f}^{12}5\mathrm{d}6\mathrm{s}^{2}$ and $4\mathrm{f}^{13}6\mathrm{s}6\mathrm{p}$ configurations implicated in Tm's visible emission and comparing their spectral behavior across trapping environments.

Understanding these higher-lying states is also important for optical readout of stored information, where fast population cycling is desirable.
Site I has the narrowest absorption features, is thermally stable, and produces strong 560~nm emission.
The ${}^{2}\mathrm{F}_{5/2}$ state at this site has a measured lifetime of approximately 30~ms \cite{marfey2025}, which limits the photon emission rate when detection relies on excitation and spontaneous decay on the 1140~nm transition.
Excitation through infrared resonances followed by upconversion pumping that yields visible emission from higher-lying states offers a route to faster readout of ${}^{2}\mathrm{F}_{5/2}$ population.
Upconversion following excitation on a narrow transition within the $4\mathrm{f}$ configuration has enabled fluorescence detection of individual Pr$^{3+}$ ions in YAG \cite{kolesov2012}.
The usefulness of these channels depends on how population reaches the emitting states and whether it returns to states that can be excited again.
Population transfer into metastable levels and recovery from these levels are therefore central to understanding the visible emission process.
Time-dependent measurements are needed to connect this emission with the population of intermediate and metastable states.

In this paper, we investigate the visible and near-infrared (NIR) emission of thulium in solid argon using excitation-emission spectroscopy and time-dependent population measurements.
We resolve six distinct excitation profiles by analyzing their visible emission patterns and compare the associated emission bands with free-atom transition energies to assess configuration-dependent matrix shifts.
We then examine the population dynamics of Site I.
Measurements of population transfer out of the ${}^{2}\mathrm{F}_{5/2}$ state, redistribution between visible-emitting states, and accumulation in long-lived levels are combined with spectroscopy of the metastable states.

\begin{figure*}[t]
    \includegraphics[width = \textwidth]{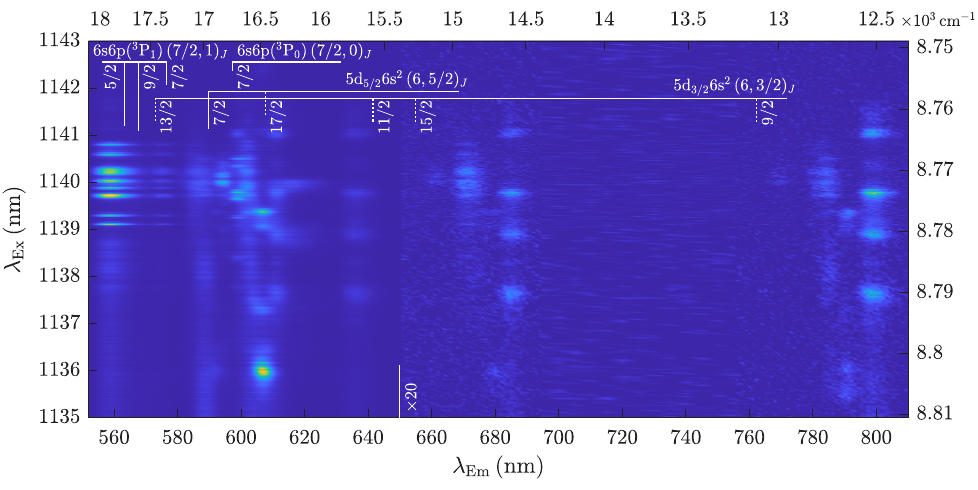}
    \caption{EEM of laser-ablated thulium in solid argon. Data at each emission wavelength $\lambda_{\mathrm{Em}}$ has been scaled by the square root of its integrated strength across $\lambda_{\mathrm{Ex}}$ to more clearly show weak features while retaining absorption profile shapes. Data collected for $\lambda_{\mathrm{Em}}>\qty{650}{\nano\meter}$ has been scaled by an additional factor of 20. Vacuum energies of low-lying $4\mathrm{f}^{13}6\mathrm{s}6\mathrm{p}$ and $4\mathrm{f}^{13}5\mathrm{d}6\mathrm{s}^2$ electronic configurations are displayed at the top of the figure, with solid lines indicating non-zero E1 transitions to the ground state and their lengths proportional to the transition strength. Dashed lines indicate vacuum-forbidden E1 transitions or when observed transition strengths are not available.}
    \label{fig:2demission}
\end{figure*}

\section{Results}
\subsection{Excitation-Emission Spectroscopy}
Inner-shell $\mathrm{f}\leftrightarrow\mathrm{f}$ transitions are minimally shifted in a rare gas matrix, allowing excitation of many trapping sites by a laser operating close to the vacuum transition energy.
To identify multiple trapping sites, an external cavity diode laser (ECDL) is scanned across the range $\lambda_{\mathrm{Ex}} =$ \qtyrange{1135}{1143}{\nano\meter} containing the magnetic dipole $\Fseven\leftrightarrow\Ffive$ transition energy with a vacuum wavelength of \qty{1140.089}{\nano\meter}.
Further excitation from the $\Ffive$ level by the same beam results in visible emission collected in the range $\lambda_{\mathrm{Em}} = $ \qtyrange{552}{810}{\nano\meter} by a grating monochromator (Acton SpectraPro-150, resolution \qty{1}{\nano\meter}).
For select emission bands, time-resolved fluorescence measurements are used to determine lifetimes.

The resulting excitation-emission matrix (EEM) is shown in Figure~\ref{fig:2demission}.
Minimal emission bands are observed at wavelengths below \qty{560}{\nano\meter} and above \qty{810}{\nano\meter} up to $\lambda_{\mathrm{Ex}}$, and the lower level of all observed transitions is assumed to be the ground state.
This assumption is justified by the fact that, due to the significant transition energies, any transition with a lower level above the ground state would originate from a high-energy region where non-radiative lifetimes are short and the spectrum is dense.
The absorption profiles of the ground state transition are inferred from the resulting visible emission, rather than from direct detection of the ${}^2\mathrm{F}_{5/2}$ decay.
In total, we identify 6 absorption profiles with narrow features from \qtyrange{1136}{1142}{\nano\meter}, each associated with a different trapping site, although it cannot be completely ruled out that some lines arise from ions or clusters.
Figure~\ref{fig:site_emission_detail} shows enlarged EEMs near \qty{585}{\nano\meter} and \qty{785}{\nano\meter} on a common excitation axis.
The emission bands near \qty{585}{\nano\meter} are tightly grouped and partially overlap.
Arrows mark selected excitation resonances of Sites I, II, and IV to clarify their overlapping contributions in this emission region.
Assignments of emission lines to vacuum configurations is performed by analysis of the EEM in two emission regions:

\begin{figure*}
    \centering
    \includegraphics[width = \textwidth]{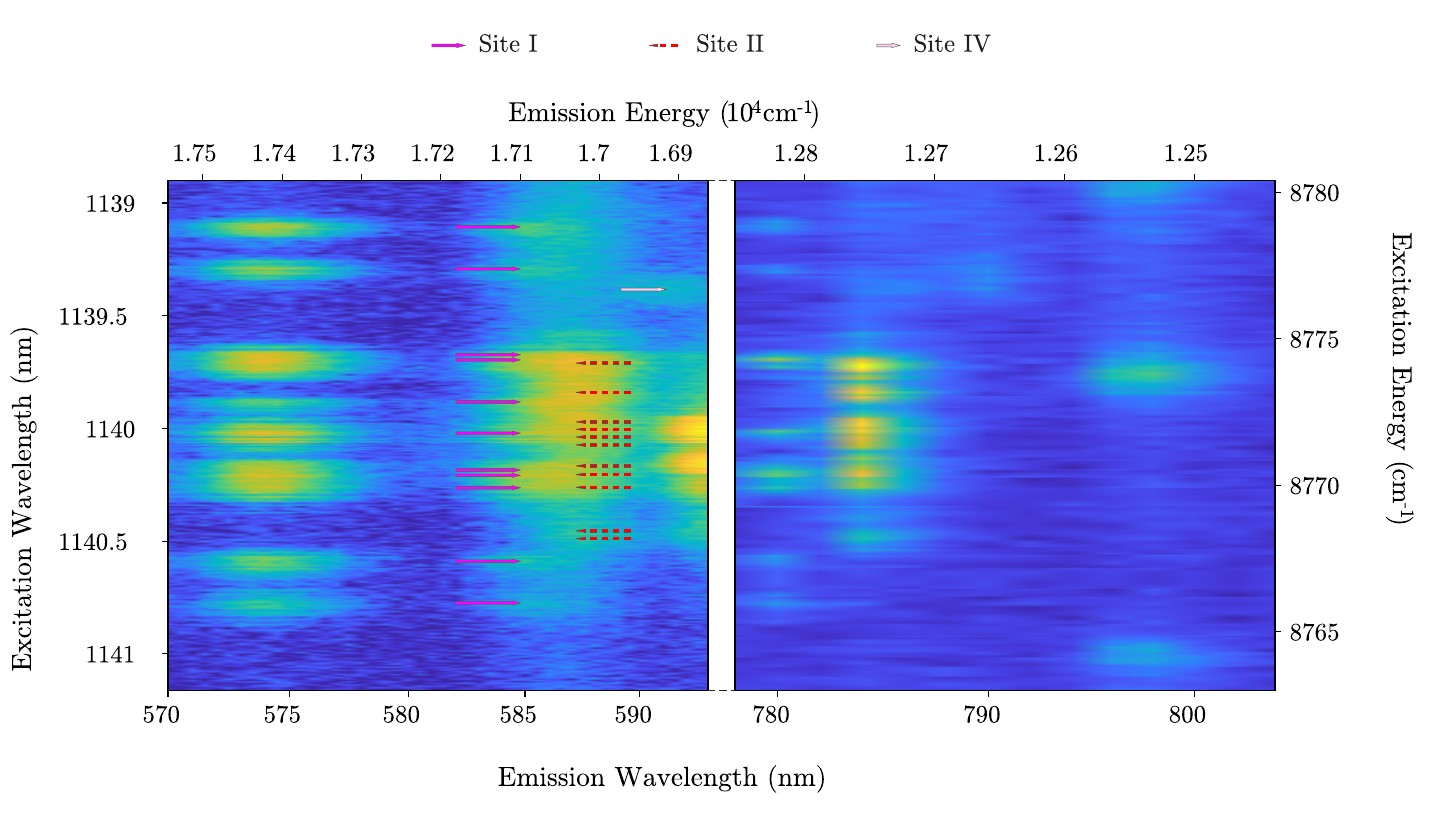}
    \caption{Excitation--emission maps of Tm in solid Ar for emission wavelengths of \qtyrange{570}{593}{\nano\meter} (left) and \qtyrange{778}{804}{\nano\meter} (right). Both panels share the excitation axis near the $\Fseven\leftrightarrow\Ffive$ transition at \qty{1140}{\nano\meter}. Arrows mark selected excitation resonances associated with visible emission from Sites I, II, and IV. The color scale is set separately for each panel.}
    \label{fig:site_emission_detail}
\end{figure*}

\textbf{650 - 810 nm.}
Five sites have emission bands grouped near \qty{670}{\nano\meter} and \qty{785}{\nano\meter}, which have been scaled additionally for clarity in Figure~\ref{fig:2demission}.
Emission in these bands most clearly distinguishes the trapping sites, and a more detailed EEM of the \qty{785}{\nano\meter} region is provided in Figure~\ref{fig:site_emission_detail}.
Since, as argued above, the ground state participates in all transitions, emission in the \qty{785}{\nano\meter} region must be assigned to the fairly isolated E1 transition from $4\mathrm{f}^{12}({}^3\mathrm{H}_6)5\mathrm{d}_{3/2}6\mathrm{s}^2\,(6,3/2)_{9/2}$ with a vacuum wavelength of \qty{762}{\nano\meter} and matrix red shifts of \qtyrange{120}{600}{\wavenumber}.
This E1 transition is only weakly forbidden, however, with an expected vacuum lifetime of \qty{2.7}{\milli\second}~\cite{li2020}, and is not likely to explain the observed lifetimes, which can be as long as \qty{100\pm15}{\milli\second} on Site I.
Lifetime increases up to a factor of 3 have been observed for other $\mathrm{f}\rightarrow\mathrm{d}$ transitions of Tm in Ne~\cite{belyaeva1986}, but extrapolating these results to the argon host requires an additional tenfold increase in the lifetime.
This trend also runs counter to the expected scaling of E1 transition strengths with the polarizability of the host~\cite{gorller-walrand1998}.
Therefore, we determine that this long lifetime may reflect a precursor state that decays by non-radiative or very weak radiative means.

Emission bands in the \qty{671}{\nano\meter} region lie closest to the vacuum energies of those states in the same configuration $4\mathrm{f}^{12}5\mathrm{d}\, (6,3/2)_J$ with $J=15/2$ and $J=11/2$ as shown in Figure~\ref{fig:2demission}, with a similar progression of red shifts across trapping sites.
We prefer an assignment of these bands as arising from the $J=11/2$ state with red shifts of \qtyrange{450}{995}{\wavenumber}, for two reasons:
First, transitions with $\Delta J = 2$, which are electric-dipole forbidden in free atoms, can become observable in rare-gas matrices through crystal-field-induced state mixing. For example, the $6s6p\,{}^3P_2 \rightarrow 6s^2\,{}^1S_0$ emission of Yb in solid Ne was attributed to Stark-like mixing induced by the matrix crystal field~\cite{lambo2021}.

Second, retaining the ordering of the vacuum $J$ levels allows the $J=15/2$ to act as the precursor state for emission at \qty{780}{\nano\meter} as described above.
On Site I, the measured lifetime of the \qty{671}{\nano\meter} emission is \qty{45\pm5}{\milli\second}.
With no estimate of the $\Delta J = 2$ transition amplitude, this lifetime may be a combination of nonradiative decay from upper levels, radiative decay into the ground state, and nonradiative decay into lower levels.

\textbf{550 - 650 nm.} Assignment of emission bands in the \qtyrange{550}{650}{\nano\meter} region is more challenging.
This region may contain multiple vacuum lines originating from unshielded configurations $4\mathrm{f}^{13}6\mathrm{s}6\mathrm{p} ({}^3\mathrm{P}_{0,1})$ and a single E1 allowed $\mathrm{f}\rightarrow\mathrm{d}$ transition from $4\mathrm{f}^{12}({}^3\mathrm{H}_6)5\mathrm{d}_{5/2}6\mathrm{s}^2 (6,5/2)_{7/2}$ with a vacuum wavelength of \qty{589.7}{\nano\meter}.
Considering stronger matrix effects on the unshielded $\mathrm{s}\rightarrow\mathrm{p}$ transitions compared to $\mathrm{f}\rightarrow\mathrm{d}$ transitions, both the absolute and relative shifts of these lines between trapping sites can be large.
For example, for multiple trapping sites of Eu in solid argon~\cite{byrne2011}, a roughly \qty{200}{\wavenumber} shift of $\mathrm{f} \rightarrow \mathrm{d}$ lines between trapping sites is associated with about a \qty{600}{\wavenumber} difference in energies for $\mathrm{s}\rightarrow\mathrm{p}$ lines.
Absolute shifts of $\mathrm{s}\rightarrow\mathrm{p}$ transitions from their vacuum values can be an order of magnitude larger than shifts for $\mathrm{f}\rightarrow\mathrm{d}$ transitions~\cite{byrne2011,belyaeva1986}.
Using this rough estimate, we expect differential shifts of $\mathrm{s}\rightarrow\mathrm{p}$ transitions across trapping sites on the order of \qty{1500}{\wavenumber}. 

\textbf{EEM Factorization and Site Specification} With this in mind, and to better isolate the sites involved in the 2D spectrum in both regions, we perform a non-negative factorization of the EEM.
Under assumptions of linearity in absorption and emission, the spectrum is decomposed into rank-6 non-negative factors
\begin{equation}
    S \approx GHK^{T},
    \label{eq:factorization}
\end{equation}
where columns of $G$ and rows of $H K^T$ are the absorption and emission profiles, respectively.
Columns of $K$ contain individual Gaussian features fitted to the original spectrum, while the matrix $H$ contains the site-feature weights.
The least-squares residual fitting is performed using the same weights as shown in Figure~\ref{fig:2demission}.
Spurious terms in $H$ are removed by associating each emission feature with a single trapping site.
Only 1 non-zero element in each column of $H$ is kept based on that element's contribution to the least-squares residual, and the absorption profiles $G$ are recalculated.
The resulting site-specific absorption and emission spectra are shown in Figure~\ref{fig:site_spectra}.
The emission spectra in Figure~\ref{fig:site_spectra} show the sparsified gaussian profiles $HK^T$ used to recompute $G$, along with the nonnegative solution $X$ of $S = GX$ determined from the recomputed $G$.
Arrows indicate features discussed below.
The sparsification procedure has the benefit of removing fitting artifacts arising from noisy measurements or non-linear dependencies on absorption profiles, but fails to identify some weak or overlapped lines.
Equipped with the absorption/emission profiles, we discuss each trapping site individually:

\begin{figure}
    \includegraphics[width = \linewidth]{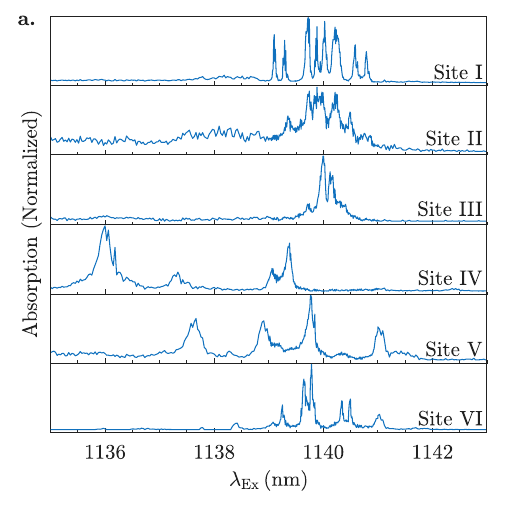}
    \includegraphics[width = \linewidth]{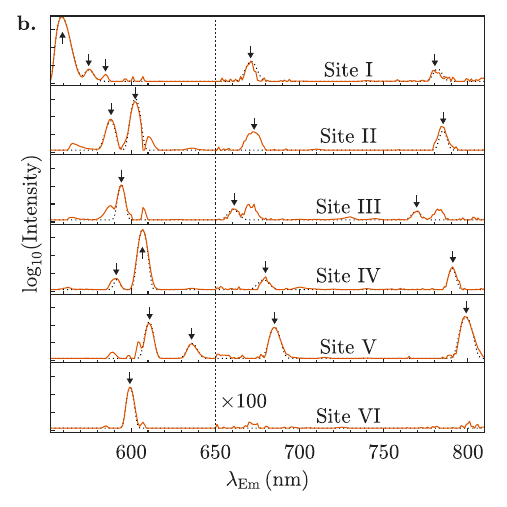} 
    \caption{Non-negative rank-6 factorization and sparsification of the excitation-emission spectrum $S = GHK^T$. \textbf{a.} Normalized absorption profiles $G$ of the six observed sites. \textbf{b.} Emission profiles of the six sites shown as the non-negative solutions $S\backslash G$ (filled lines), and the sparsified $HK^T$ (dotted lines). The emission spectra are plotted on a log scale with additional scaling at $\lambda_{\mathrm{Em}}>650\,\mathrm{nm}$ to highlight weaker features.}
    \label{fig:site_spectra}
\end{figure}

\textbf{Site I} has the narrowest absorption features in the range \qtyrange{1139.1}{1140.8}{\nano\meter} with a weak sideband roughly \qty{15}{\wavenumber} from its mean transition energy.
The crystal field and hyperfine structure for Site I has been previously resolved~\cite{marfey2025,gaire2023}.
Site I produces distinctly strong emission at \qty{560}{\nano\meter} with a lifetime of \qty{600}{\nano\second} and a weak emission line at \qty{575}{\nano\meter} whose lifetime is faster than the AOM turn-off time (\qty{150}{\nano\second}).
This short lifetime is attributed to non-radiative decay into nearby states, however the radiative lifetime is likely competingly short considering the line is observed at all.
Thus, both of these lines are attributed to strong $\mathrm{s}\leftrightarrow\mathrm{p}$ transitions with matrix blue shifts on the order of \qty{500}{\wavenumber}.
With the assumption that the term ordering is unchanged from the vacuum levels, we assign the lower energy emission band at \qty{575}{\nano\meter} to the ${}^3\mathrm{P}_0\,(7/2,0)_{7/2}$ state and the higher energy band at \qty{560}{\nano\meter} to the ${}^3\mathrm{P}_1$ manifold.
Within the upper manifold, we assign the band to the lowest $J=7/2$, assuming the emission from the higher lying $J=9/2$ and $J= 5/2$ states is rapidly quenched by internal relaxation.
Apart from these $\mathrm{s}\rightarrow\mathrm{p}$ bands, we observe longer-lived emission features from Site I at \qty{585}{\nano\meter}, \qty{671}{\nano\meter}, and \qty{780}{\nano\meter}, with lifetimes of \qtylist{4.3;45;100}{\milli\second}, respectively.
The latter two are assigned to $\mathrm{f}\rightarrow\mathrm{d}$ transitions by previous reasoning.
Taking the \qty{}{\milli\second}-scale lifetimes as characteristic of the $\mathrm{f}\rightarrow\mathrm{d}$ transitions due to internal relaxation within the $\mathrm{f}^{12}\mathrm{d}$ configuration, we assign the remaining feature at \qty{585}{\nano\meter} to the $5\mathrm{d}_{5/2}6\mathrm{s}^2\,(6,5/2)_{7/2}$ state with a matrix blue shift of \qty{136}{\wavenumber}.
Details surrounding the excitation dynamics of Site I are presented in a later section.

The reasoning and assignments made for Site I provide a framework that can be judiciously extended to Sites II-VI, though a site-by-site analysis is still warranted.

\textbf{Site II} has absorption features in the range \qtyrange{1139.7}{1140.5}{\nano\meter} and a weak sideband similarly spaced at \qty{15}{\wavenumber}.
The crystal field level structure of Site II has also been determined~\cite{marfey2025}.
Site II also features weak emission bands at \qty{588}{\nano\meter}, \qty{673}{\nano\meter}, and \qty{784}{\nano\meter}, with lifetimes similar to those of Site I.
The latter are assigned above and the band at \qty{588}{\nano\meter} is assigned to the $5\mathrm{d}_{5/2}6\mathrm{s}^2\,(6,5/2)_{7/2}$ state, reflecting small deviations from the structure of Site I for $\mathrm{f}\rightarrow\mathrm{d}$ transitions.
An additional band at \qty{603}{\nano\meter} is observed for Site II, close to the vacuum transition energy of the $({}^3\mathrm{P}_0)\,(7/2,0)_{7/2}$ state at \qty{597.29}{\nano\meter}.

\textbf{Site III} has the narrowest total spread of absorption, centered around \qty{1140.05}{\nano\meter}.
This site features correspondingly small red shifts for its $\mathrm{f}\rightarrow\mathrm{d}$ lines at \qty{661}{\nano\meter} and \qty{770}{\nano\meter}.
The remaining feature at \qty{594}{\nano\meter} remains unassigned.
It is interesting to note, however, that assignment of this band to the nearest $(6,5/2)_{7/2}$ state and reassignment of the \qty{661}{\nano\meter} band to $(6,3/2)_{15/2}$ would indicate a near-scalar red shift of these $\mathrm{f}^{12}\mathrm{d}$ lines by \qtyrange{120}{140}{\wavenumber}.
We also note here the difficulty in separating Sites II and III by the non-negative factorization (Figure~\ref{fig:site_spectra}) due to their weak emission features and significant overlap in excitation.

\textbf{Site IV} has its strongest absorption feature at \qty{1136}{\nano\meter} and a large deviation of its barycenter from the vacuum $\Fseven\rightarrow\Ffive$ transition at \qty{1140.089}{\nano\meter}.
Determination of the true average transition energy by crystal field analysis is made difficult by the small number of features and the large splitting, on the order of tens of \qty{}{\wavenumber}.
Site IV has $\mathrm{f}\rightarrow\mathrm{d}$ transitions at \qty{591}{\nano\meter}, \qty{680}{\nano\meter}, and \qty{791}{\nano\meter}, assigned to the $(6,5/2)_{7/2}$, $(6,3/2)_{11/2}$, and $(6,3/2)_{9/2}$ states.
The strong emission band at \qty{607}{\nano\meter} is likely an $\mathrm{s}\rightarrow\mathrm{p}$ transition.

\textbf{Site V} has similarly sparse absorption features from \qtyrange{1137.7}{1141}{\nano\meter} and the largest red shifts of its $\mathrm{f}\rightarrow\mathrm{d}$ emission bands at \qty{685}{\nano\meter} and \qty{799}{\nano\meter}.
This is accompanied by two bands at \qty{611}{\nano\meter} and \qty{636}{\nano\meter}, which remain unassigned but share a similar spacing to the vacuum $({}^3\mathrm{P}_0)\,(7/2, 0)_{7/2}$ and $({}^3\mathrm{P}_1)\,(7/2,1)_{7/2}$ energies.

\textbf{Site VI} has absorption features from \qtyrange{1138.3}{1141}{\nano\meter} and its sole emission feature at \qty{600}{\nano\meter}.
This band remains unassigned, and we note that the presence of this site is not reproduced across varying growth conditions.

Holistically, many sites have weak emission features tightly grouped around \qty{588}{\nano\meter}, \qty{670}{\nano\meter}, and \qty{785}{\nano\meter}, characteristic of $\mathrm{f}\rightarrow\mathrm{d}$ transitions.
As expected, the positions of the stronger bands vary widely across trapping sites, characteristic of $\mathrm{s}\rightarrow\mathrm{p}$ transitions.
However, specific assignements to any sublevel of the ${}^3\mathrm{P}_0$ or ${}^3\mathrm{P}_1$ configurations are not clear.

\section{Site I}
Absorption and emission profiles for Site I indicate high count rates for the \qty{560}{\nano\meter} transition paired with narrow absorption lines on the ground state transition.
Additionally, this site is thermally stable upon anneals to \qty{30}{\kelvin} and appears consistently across multiple growth cycles.
This prompts further exploration into the excitation dynamics of site I, where a \qty{560}{\nano\meter} narrow bandpass filter is used to isolate processes involving this site.
We note that \qty{560}{\nano\meter} emission requires a $\geq 3$-step absorption process from the ground state $\Fseven$, where the first step is the spin-orbit transition to the metastable $\Ffive$ level with a lifetime of \qty{30}{\milli\second}.
We identify several transitions, including some originating from the $\Ffive$ level, which contribute to the observed emission at \qty{560}{\nano\meter}.

\subsection{$\Ffive$ Absorption}
\begin{figure}
    \includegraphics[width = \linewidth]{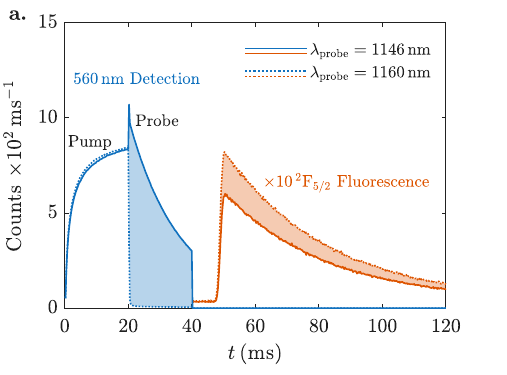}
    \includegraphics[width = \linewidth]{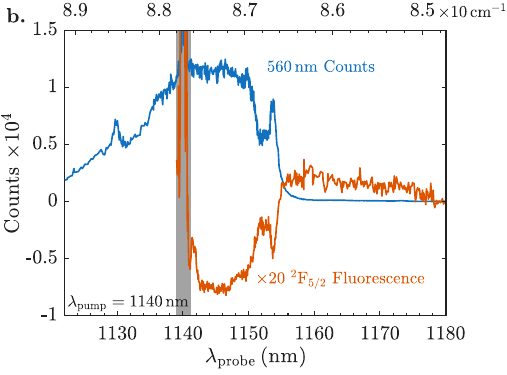}
    \caption{Removal of $\Ffive$ population by secondary absorption with a minimum energy corresponding to \qty{1153}{\nano\meter}. \textbf{a.} Averaged experiment traces near \qty{1146}{\nano\meter} and \qty{1160}{\nano\meter}. \qty{560}{\nano\meter} counts are collected during the pump and probe, while long-lived $\Ffive$ fluorescence is collected after a short delay. The highlighted regions indicate the change in signal between the two selected probe wavelengths. \textbf{b.} Integrated \qty{560}{\nano\meter} and $\Ffive$ counts for each choice of $\lambda_{\mathrm{probe}}$. The baseline is chosen such that the integrated counts at $\lambda_{\mathrm{probe}} = \qty{1180}{\nano\meter}$ are zero. Near and below \qty{1140}{\nano\meter}, the $\Ffive$ fluorescence is dominated by reabsorption from the probe beam. This data has been omitted.}
    \label{fig:5/2_depump}
\end{figure}

A pump laser fixed on a fairly isolated Site I absorption line at \qty{1140.585}{\nano\meter} initializes population in the metastable $\Ffive$ and higher levels involved in the \qty{560}{\nano\meter} emission process.
A probe beam at $\lambda_{\mathrm{probe}}$ following the pump results in further \qty{560}{\nano\meter} emission and a relative decrease in $\Ffive$ population, which is measured as a decrease in the \qty{30}{\milli\second}-lived $\Ffive$ fluorescence after both pulses are complete.
Experimental traces for two choices of $\lambda_{\mathrm{probe}}$ are shown in Figure~\ref{fig:5/2_depump}a.
Figure~\ref{fig:5/2_depump}b shows the correlation between \qty{560}{\nano\meter} counts and a decreased $\Ffive$ population as the probe is scanned from \qtyrange{1120}{1180}{\nano\meter}.

The spectrum in Figure~\ref{fig:5/2_depump} is broad with a sharper feature on the red edge at \qty{1153}{\nano\meter}, corresponding to the minimum energy that is required to bring atoms out of the metastable $\Ffive$ level back to the ground state through the \qty{560}{\nano\meter} channel.
A possible explanation for this spectrum is a group of closely spaced levels (such as crystal field levels), with each one having a phonon side band, such that only the zero-phonon line of the reddest feature remains sharp.
The proximity of this resonance to the ground state transition is entirely coincidental, and a similar transition has not yet been observed for other sites.
In combination with the ground state transition, the 2-absorption process results in a total energy closely matching the short-lived emission line observed at \qty{575}{\nano\meter}.
The decay of \qty{575}{\nano\meter} emission is faster than the $~\qty{150}{\nano\second}$ turn-off time of the optical components, which we attribute to rapid non-radiative decay into lower-lying metastable levels, including the \qty{4.3}{\milli\second}-lived state emitting at \qty{585}{\nano\meter}.
At an elevated temperature of \qty{20}{\kelvin}, the \qty{575}{\nano\meter} emission line cannot be observed, which we attribute to an even faster non-radiative decay outcompeting the radiative process completely.

\subsection{\qty{585}{\nano\meter} State Absorption}
\begin{figure}
    \centering
    \includegraphics[width = \linewidth]{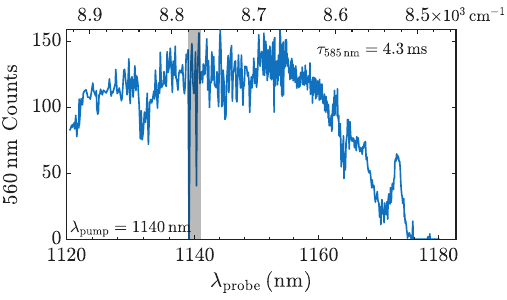}

    \caption{Lineshape of excitation from the \qty{585}{\nano\meter} emitting level. The data are taken as the component matching the \qty{585}{\nano\meter} lifetime of \qty{4.3}{\milli\second} in the experiment described in the text. }
    \label{fig:585_depump}
\end{figure}

Despite excitation of $\Ffive$ population by the \qty{1153}{\nano\meter} resonance, a tertiary transition is necessary to reach the energy required for \qty{560}{\nano\meter} emission.
We identify one such transition as an excitation out of the \qty{585}{\nano\meter} emitting state, whose measured lifetime is \qty{4.3}{\milli\second}.
To identify \qty{560}{\nano\meter} light caused by this transition, we introduce a variable delay between the pump and probe;
as a function of the delay, the amount of \qty{560}{\nano\meter} light is determined by the natural decay of the \qty{585}{\nano\meter} emitting state and the $\Ffive$ state.
For a probe with wavelength between \qtylist{1153;1172}{\nano\meter}, the signal mostly decays within \qty{4.3}{\milli\second}, as the \qty{585}{\nano\meter} state depletes naturally.
For light with wavelength below \qty{1153}{\nano\meter}, there is a component decaying at \qty{4.3}{\milli\second}, but also a long-lived component associated with refilling the \qty{585}{\nano\meter} state from the $\Ffive$ level, as shown in Figure~\ref{fig:5/2_depump}.
By fitting \qty{560}{\nano\meter} counts to a double-exponential function of the delay, we extract the component which decays with \qty{4.3}{\milli\second} lifetime as the efficacy of light at pumping the \qty{585}{\nano\meter} level, resulting in the spectrum shown in Figure~\ref{fig:585_depump}.

The excitation feature in Figure~\ref{fig:585_depump} is even broader than that in Figure~\ref{fig:5/2_depump}, but can also be explained by a zero-phonon feature at \qty{1172}{\nano\meter} and a broad phonon wing overlapping with other features.
After excitation by this transition, the energy far exceeds that required for emission at \qty{560}{\nano\meter}, and even surpasses the energy of the strong cooling transitions at and above \qty{410}{\nano\meter}~\cite{kolachevsky2007}.
However, we observe no strong emission peaks in the \qtyrange{400}{550}{\nano\meter} range, indicating rapid relaxation into the state emitting at \qty{560}{\nano\meter} following the 3-stage excitation.
Quenching of strong vacuum transitions is not uncommon and has been observed for the ${}^1\mathrm{P}_1$ state of Yb~\cite{kleshchina2019,xu2011}, especially in rare gases heavier than neon.
This further supports the notion that radiative decays from high lying excitations into intermediate levels are not a likely explanation for observed emission features.

\subsection{$\mathrm{f}^{12}\mathrm{d}$ Absorption}
\begin{figure}
    \centering
    \includegraphics[width = \linewidth]{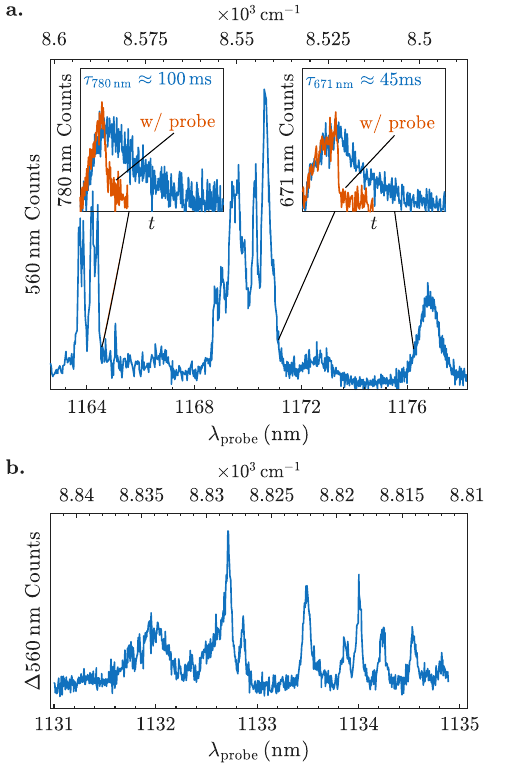}
    \caption{ \textbf{a.} Depletion of long-lived $\mathrm{f}^{12}\mathrm{d}$ levels emitting at \qty{670}{\nano\meter} and \qty{780}{\nano\meter} by $\lambda_{\mathrm{probe}}$, measured through the \qty{560}{\nano\meter} decay channel. The insets indicate the reduction in \qty{670}{\nano\meter} and \qty{780}{\nano\meter} emission during the probe. \textbf{b.} Depletion lines near $\lambda_{\mathrm{probe}} = \qty{1133}{\nano\meter}$. Overlap with the $\Ffive$ excitation has been subtracted, and the corresponding reduction in emission occurs at \qty{670}{\nano\meter}.}
    \label{fig:metastable_depump}
\end{figure}

We investigate further the emission bands near \qty{670}{\nano\meter} and \qty{785}{\nano\meter}, as seen from the EEM in Figure~\ref{fig:2demission} and the decomposition in Figure~\ref{fig:site_spectra}.
Sites I-V have long lived emission bands in these regions originating from the $\mathrm{f}^{12}\mathrm{d}$ configuration.
For Site I, the bands at \qty{671}{\nano\meter} and \qty{780}{\nano\meter} have lifetimes of \qty{45\pm5}{\milli\second} and \qty{100\pm15}{\milli\second}, longer than the \qty{30.5}{\milli\second} lifetime of the $\Fseven \leftrightarrow \Ffive$ inner-shell transition.
Following an initialization of these states and a delay to allow natural depletion of any short-lived state, reabsorption of these long-lived states by a probe beam results in emission at \qty{560}{\nano\meter}.
For $\lambda_{\mathrm{probe}}>\qty{1153}{\nano\meter}$, the absorption lines are measured directly by the \qty{560}{\nano\meter} luminescence and are shown in Figure~\ref{fig:metastable_depump}.
For $\lambda_{\mathrm{probe}}<\qty{1153}{\nano\meter}$, background counts due to excitation of remaining $\Ffive$ population are subtracted from the spectrum.

We observe four narrow absorption features in our laser-accessible range of $\lambda_{\mathrm{probe}}=\qtyrange{1120}{1180}{\nano\meter}$ at \qty{1133}{\nano\meter}, \qty{1164}{\nano\meter}, \qty{1170}{\nano\meter}, and \qty{1177}{\nano\meter}.
The multiplet structure of the lines at \qtylist{1133;1164;1169}{\nano\meter} which span \qtyrange{10}{25}{\wavenumber} are attributed to crystal field splitting within the $4\mathrm{f}^{12}5\mathrm{d}$ configuration.
The narrow linewidths $<\qty{1}{\nano\meter}$ indicate a possible inner-shell transition within this excited configuration.

\begin{figure}
    \centering
    \includegraphics{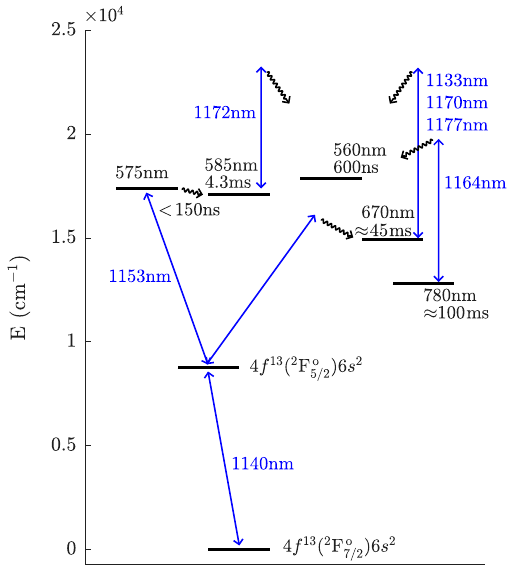}
    \caption{Site I energy level structure with relevant transitions}
    \label{fig:SiteIFull}
\end{figure}

We measure a rapid decrease of either the \qty{671}{\nano\meter} or \qty{780}{\nano\meter} emission under influence of the probe beam (shown as an inset in Figure~\ref{fig:metastable_depump}) to infer the lower level of each absorption line.
The three sets of lines at \qtylist{1133;1169;1177}{\nano\meter} deplete the state emitting at \qty{671}{\nano\meter}, while the lines at \qty{1164}{\nano\meter} deplete the state emitting at \qty{780}{\nano\meter}.
Owing to the discussion of the \qty{671}{\nano\meter} and \qty{780}{\nano\meter} lifetimes above, the energy of the lower level cannot be definitively established, if the long lifetimes are due to internal relaxation.
This uncertainty hinders attempts at crystal-field analysis of these lines, which could help establish the angular momenta $J$ of the upper and lower manifolds.
Figure~\ref{fig:SiteIFull} summarizes the absorption and emission processes observed on Site I.

\section{Discussion}

The excitation-emission spectroscopy identifies 6 absorption-emission profiles for argon-isolated thulium atoms, which are attributed to multiple trapping sites in the matrix.
This contrasts with the few stable trapping sites of high symmetry predicted for similarly sized atoms~\cite{ozerov2019} and previous measurements of thulium atoms in solid Ne~\cite{belyaeva1986} where all observed lines were attributed to a single site.
Laser ablation as the source of atoms may play a role in the large number of sites observed, as it has been shown that laser ablated species can occupy novel sites in the matrix, inducing large variations in matrix shifts, especially for $\mathrm{s}\rightarrow\mathrm{p}$ transitions~\cite{fajardo1991,tam1993}.
Singly and doubly ionized thulium have ground state transitions at \qty{1140.3}{\nano\meter} and \qty{1139.73}{\nano\meter}, respectively, so the presence of ionic species cannot entirely be ruled out.
The exceptions are both Site I and II where crystal field analysis has shown very close agreement of the average transition energy to that of neutral thulium atoms~\cite{marfey2025}.
The lack of broad visible bands and vibrational progressions for clusters and dimers of similar atoms~\cite{klotzbucher1997,suzer1988} indicate that these types of ablation products are not present.

For Site I in particular, the emission at \qty{560}{\nano\meter} is partially due to reexcitation of metastable levels, some of which are populated by non-radiative decay.
We find two transitions which coincidentally overlap with the ground state transition, producing more than enough energy to produce the emission band at \qty{560}{\nano\meter} following significant non-radiative decay.
Additionally, narrow absorption lines displaying crystal field splitting recycle atoms shelved in the metastable $4\mathrm{f}^{12}5\mathrm{d}$ states responsible for emission at \qty{670}{\nano\meter} and \qty{780}{\nano\meter}.
These states contribute to atom loss in thulium traps~\cite{sukachev2010} and their vacuum lifetimes have only been estimated~\cite{li2020}.

We observe lifetimes far exceeding those predicted, attributed to slow non-radiative decay from the nearest higher-lying state.
By comparison with with the dynamics of Er in solid Ar and Ne~\cite{zaitsev2026}, the \qty{40}{\milli\second} lifetime of the $J = 11/2$ level would arise from a multi-phonon transition across the \qty{869}{\wavenumber} gap lying above it, and the \qty{100}{\milli\second} lifetime of the $J = 9/2$ level would arise from a \qty{2152}{\wavenumber} gap.
Additionally, it would be expected that radiative transitions from $J=11/2$ would be strongly quenched by relaxation crossing the \qty{316}{\wavenumber} gap into $J=15/2$.
The energy gaps quoted here are for the vacuum levels and could be altered in the matrix.
If the \qty{670}{\nano\meter} emission was assigned to $J=15/2$, then the lifetime of the \qty{780}{\nano\meter} emission would at most match that of the \qty{670}{\nano\meter}.
Thus, the lifetime of the \qty{780}{\nano\meter} emission is unusally long regardless of the assignment of the line at \qty{670}{\nano\meter}, without enforcing special selection rules for non-radiative decays or between crystal field sublevels.
Laser excitation of the $\mathrm{f}^{12}\mathrm{d}$ lines would establish the source of these long lifetimes and aide in determining the $\mathrm{d}$-shell crystal field splitting.

\section{Conclusion}
We have identified multiple trapping sites of thulium in solid argon using excitation-emission spectroscopy, and established lifetimes and secondary excitation processes on Site I.
The overall structure of Site I is shown in Figure~\ref{fig:SiteIFull} along with the specific excitation features, emission features, and lifetimes observed.

\section{Acknowledgements}
This material is based upon work supported by the U.S. National Science Foundation under Grant No. PHY-2310394. J.M. acknowledges support from the NSF Graduate Research Fellowship Program under Grant No. SFT-2545526.

\bibliography{references}

@article{sukachev2010,
	title = {Magneto-optical trap for thulium atoms},
	volume = {82},
	url = {https://link.aps.org/doi/10.1103/PhysRevA.82.011405},
	doi = {10.1103/PhysRevA.82.011405},
	number = {1},
	urldate = {2025-11-26},
	journal = {Physical Review A},
	publisher = {American Physical Society},
	author = {Sukachev, D. and Sokolov, A. and Chebakov, K. and Akimov, A. and Kanorsky, S. and Kolachevsky, N. and Sorokin, V.},
	month = jul,
	year = {2010},
	pages = {011405},
}

@article{li2020,
	title = {Theoretical study of the spectroscopic properties of mendelevium ({Z}=101{\textless}math{\textgreater}{\textless}mrow is="true"{\textgreater}{\textless}mi is="true"{\textgreater}{Z}{\textless}/mi{\textgreater}{\textless}mo linebreak="goodbreak" is="true"{\textgreater}={\textless}/mo{\textgreater}{\textless}mn is="true"{\textgreater}101{\textless}/mn{\textgreater}{\textless}/mrow{\textgreater}{\textless}/math{\textgreater})},
	volume = {247},
	issn = {0022-4073},
	url = {https://www.sciencedirect.com/science/article/pii/S0022407320301217},
	doi = {10.1016/j.jqsrt.2020.106943},
	urldate = {2026-07-01},
	journal = {Journal of Quantitative Spectroscopy and Radiative Transfer},
	author = {Li, Jiguang and Dzuba, Vladimir},
	month = may,
	year = {2020},
	pages = {106943},
}

@article{belyaeva1986,
	title = {Effect of the environment on the spectral and temporal characteristics of thulium atoms isolated in solid neon},
	volume = {60},
	journal = {Opt. Spektrosk.},
	author = {Belyaeva, A. A. and Predtechenskii, Yu. B.},
	month = jun,
	year = {1986},
	pages = {1130--1137},
}

@article{kleshchina2019,
	title = {Triplet emission of atomic ytterbium isolated in a xenon matrix},
	volume = {45},
	issn = {1063-777X},
	doi = {10.1063/1.5111293},
	number = {7},
	journal = {Low Temperature Physics},
	author = {Kleshchina, N. N. and Kalinina, I. S. and Lambo, R. and Buchachenko, A. A. and Bezrukov, D. S. and Hu, S.-M.},
	year = {2019},
	pages = {707--714},
}

@article{lambo2012,
	title = {Electronic spectroscopy of ytterbium in a neon matrix},
	volume = {137},
	issn = {0021-9606},
	doi = {10.1063/1.4768419},
	number = {20},
	journal = {The Journal of Chemical Physics},
	author = {Lambo, R. and Buchachenko, A. A. and Wu, L. and Tan, Y. and Wang, J. and Sun, Y. R. and Liu, A.-W. and Hu, S.-M.},
	year = {2012},
}

@article{lambo2021,
	title = {High-resolution spectroscopy of neutral {Yb} atoms in a solid {Ne} matrix},
	volume = {104},
	issn = {2469-9926},
	doi = {10.1103/physreva.104.062809},
	number = {6},
	journal = {Physical Review A},
	author = {Lambo, R. and Xu, C. Y. and Pratt, S. T. and Xu, H. and Zappala, J. C. and Bailey, K. G. and Lu, Z. T. and Mueller, P. and O'Connor, T. P. and Kamorzin, B. B. and Bezrukov, D. S. and Xie, Y. Q. and Buchachenko, A. A. and Singh, J. T.},
	year = {2021},
}

@article{xu2011,
	title = {Optical {Excitation} and {Decay} {Dynamics} of {Ytterbium} {Atoms} {Embedded} in a {Solid} {Neon} {Matrix}},
	volume = {107},
	doi = {10.1103/PhysRevLett.107.093001},
	number = {9},
	journal = {Physical Review Letters},
	author = {Xu, C. Y. and Hu, S. M. and Singh, J. and Bailey, K. and Lu, Z. T. and Mueller, P. and O’Connor, T. P. and Welp, U.},
	year = {2011},
	pages = {093001},
}

@article{suzer1988,
	title = {Optical spectra of {Yb} atoms and dimers in rare gas matrices},
	volume = {89},
	issn = {0021-9606, 1089-7690},
	url = {https://pubs.aip.org/jcp/article/89/9/5514/221037/Optical-spectra-of-Yb-atoms-and-dimers-in-rare-gas},
	doi = {10.1063/1.455576},
	language = {en},
	number = {9},
	urldate = {2026-07-14},
	journal = {The Journal of Chemical Physics},
	author = {Suzer, Sefik and Andrews, Lester},
	month = nov,
	year = {1988},
	pages = {5514--5516},
}

@article{gaire2019,
  author  = {Gaire, Vinod and Raman, Chandra S. and Parker, Colin V.},
  title   = {Subnanometer optical linewidth of thulium atoms in rare-gas crystals},
  journal = {Physical Review A},
  year    = {2019},
  volume  = {99},
  number  = {2},
  pages   = {022505},
  doi     = {10.1103/PhysRevA.99.022505},
  url     = {https://doi.org/10.1103/PhysRevA.99.022505},
}

@article{gaire2023,
  author  = {Gaire, Vinod and Do, Mi Y. and Pei, Yiting and
             Semenova, Anthony and Parker, Colin V.},
  title   = {High resolution spectroscopy of thulium atoms implanted in solid noble gas crystals},
  journal = {Physical Review B},
  year    = {2023},
  volume  = {108},
  number  = {21},
  pages   = {214101},
  doi     = {10.1103/PhysRevB.108.214101},
  url     = {https://doi.org/10.1103/PhysRevB.108.214101},
}

@article{marfey2025,
  author  = {Marfey, Jason and Semenova, Anthony and Parker, Colin V.},
  title   = {Crystal fields and {Zeeman} effect for thulium in solid argon},
  journal = {Physical Review B},
  year    = {2025},
  volume  = {112},
  number  = {1},
  pages   = {014101},
  doi     = {10.1103/x5zx-ykff},
  url     = {https://doi.org/10.1103/x5zx-ykff},
}

@article{lancaster2024,
  author  = {Lancaster, David M. and Dargyte, Ugne and Weinstein, Jonathan D.},
  title   = {Optical spin readout of single rubidium atoms trapped in solid neon},
  journal = {Physical Review Research},
  year    = {2024},
  volume  = {6},
  number  = {1},
  pages   = {L012048},
  doi     = {10.1103/PhysRevResearch.6.L012048},
  url     = {https://doi.org/10.1103/PhysRevResearch.6.L012048},
}

@article{fajardo1991,
  author  = {Fajardo, Mario E. and Carrick, P. G. and Kenney, III, John W.},
  title   = {Matrix isolation spectroscopy of metal atoms generated by laser ablation. I. The {Li/Ar}, {Li/Kr}, and {Li/Xe} systems},
  journal = {The Journal of Chemical Physics},
  year    = {1991},
  volume  = {94},
  number  = {9},
  pages   = {5812--5825},
  doi     = {10.1063/1.460465},
  url     = {https://doi.org/10.1063/1.460465}
}

@article{byrne2011,
  author  = {Byrne, Owen and McCaffrey, John G.},
  title   = {{Eu/RG} absorption and excitation spectroscopy in the solid rare gases: State dependence of crystal field splitting and {Jahn--Teller} coupling},
  journal = {The Journal of Chemical Physics},
  year    = {2011},
  volume  = {134},
  number  = {12},
  pages   = {124501},
  doi     = {10.1063/1.3564947},
  url     = {https://doi.org/10.1063/1.3564947}
}

@misc{zaitsev2026,
  author        = {Zaitsev, Evgenii and Dinerstein, Alec and Zou, Shan and Peretti, Charles and Wen, Yutian and Guo, Wei and Hartland, Gregory V. and Huang, Yizhong and Jin, Dafei},
  title         = {Telecom-Band Optical Transitions of Erbium Atoms in Rare-Gas Solids},
  year          = {2026},
  month         = oct,
  eprint        = {2610.00918},
  archivePrefix = {arXiv},
  primaryClass  = {quant-ph},
  doi           = {10.48550/arXiv.2610.00918},
  url           = {https://arxiv.org/abs/2610.00918},
  note          = {Preprint}
}

@article{chambers2019,
	title = {Imaging individual barium atoms in solid xenon for barium tagging in {nEXO}},
	volume = {569},
	copyright = {2019 The Author(s), under exclusive licence to Springer Nature Limited},
	issn = {1476-4687},
	url = {https://www.nature.com/articles/s41586-019-1169-4},
	doi = {10.1038/s41586-019-1169-4},
	language = {en},
	number = {7755},
	urldate = {2026-11-04},
	journal = {Nature},
	publisher = {Nature Publishing Group},
	author = {Chambers, C. and Walton, T. and Fairbank, D. and Craycraft, A. and Yahne, D. R. and Todd, J. and Iverson, A. and Fairbank, W. and Alamre, A. and Albert, J. B. and Anton, G. and Arnquist, I. J. and Badhrees, I. and Barbeau, P. S. and Beck, D. and Belov, V. and Bhatta, T. and Bourque, F. and Brodsky, J. P. and Brown, E. and Brunner, T. and Burenkov, A. and Cao, G. F. and Cao, L. and Cen, W. R. and Charlebois, S. A. and Chiu, M. and Cleveland, B. and Coon, M. and Côté, M. and Cree, W. and Dalmasson, J. and Daniels, T. and Darroch, L. and Daugherty, S. J. and Daughhetee, J. and Delaquis, S. and Der Mesrobian-Kabakian, A. and DeVoe, R. and Dilling, J. and Ding, Y. Y. and Dolinski, M. J. and Dragone, A. and Echevers, J. and Fabris, L. and Farine, J. and Feyzbakhsh, S. and Fontaine, R. and Fudenberg, D. and Gallina, G. and Giacomini, G. and Gornea, R. and Gratta, G. and Hansen, E. V. and Heffner, M. and Hoppe, E. W. and Hößl, J. and House, A. and Hufschmidt, P. and Hughes, M. and Ito, Y. and Jamil, A. and Jessiman, C. and Jewell, M. J. and Jiang, X. S. and Karelin, A. and Kaufman, L. J. and Kodroff, D. and Koffas, T. and Kravitz, S. and Krücken, R. and Kuchenkov, A. and Kumar, K. S. and Lan, Y. and Larson, A. and Leonard, D. S. and Li, G. and Li, S. and Li, Z. and Licciardi, C. and Lin, Y. H. and Lv, P. and MacLellan, R. and Michel, T. and Mong, B. and Moore, D. C. and Murray, K. and Newby, R. J. and Ning, Z. and Njoya, O. and Nolet, F. and Nusair, O. and Odgers, K. and Odian, A. and Oriunno, M. and Orrell, J. L. and Ortega, G. S. and Ostrovskiy, I. and Overman, C. T. and Parent, S. and Piepke, A. and Pocar, A. and Pratte, J.-F. and Qiu, D. and Radeka, V. and Raguzin, E. and Rao, T. and Rescia, S. and Retière, F. and Robinson, A. and Rossignol, T. and Rowson, P. C. and Roy, N. and Saldanha, R. and Sangiorgio, S. and Schmidt, S. and Schneider, J. and Schubert, A. and Skarpaas, K. and Soma, A. K. and St-Hilaire, G. and Stekhanov, V. and Stiegler, T. and Sun, X. L. and Tarka, M. and Tolba, T. and Totev, T. I. and Tsang, R. and Tsang, T. and Vachon, F. and Veenstra, B. and Veeraraghavan, V. and Visser, G. and Vuilleumier, J.-L. and Wagenpfeil, M. and Wang, Q. and Watkins, J. and Weber, M. and Wei, W. and Wen, L. J. and Wichoski, U. and Wrede, G. and Wu, S. X. and Wu, W. H. and Xia, Q. and Yang, L. and Yen, Y.-R. and Zeldovich, O. and Zhang, X. and Zhao, J. and Zhou, Y. and Ziegler, T. and {nEXO Collaboration}},
	month = may,
	year = {2019},
	pages = {203--207},
}

@article{ishikawa1997,
	title = {Laser spectroscopy of thulium atoms implanted in liquid and solid ${}^{4}${He}},
	volume = {56},
	url = {https://link.aps.org/doi/10.1103/PhysRevB.56.780},
	doi = {10.1103/PhysRevB.56.780},
	number = {2},
	journal = {Physical Review B},
	author = {Ishikawa, Kiyoshi and Hatakeyama, Atsushi and Gosyono-o, Koichi and Wada, Shigeaki and Takahashi, Yoshiro and Yabuzaki, Tsutomu},
	month = jul,
	year = {1997},
	pages = {780--787},
}

@article{golovizin2019,
	title = {Inner-shell clock transition in atomic thulium with a small blackbody radiation shift},
	volume = {10},
	copyright = {2019 The Author(s)},
	issn = {2041-1723},
	url = {https://www.nature.com/articles/s41467-019-09706-9},
	doi = {10.1038/s41467-019-09706-9},
	language = {en},
	number = {1},
	urldate = {2025-10-28},
	journal = {Nature Communications},
	author = {Golovizin, A. and Fedorova, E. and Tregubov, D. and Sukachev, D. and Khabarova, K. and Sorokin, V. and Kolachevsky, N.},
	month = apr,
	year = {2019},
	pages = {1724},
}

@article{byrne2011luminescence,
  author  = {Byrne, Owen and McCaffrey, John G.},
  title   = {Site-selected luminescence of atomic europium in the solid rare gases},
  journal = {The Journal of Chemical Physics},
  year    = {2011},
  volume  = {135},
  number  = {2},
  pages   = {024507},
  doi     = {10.1063/1.3609116},
}

@article{kolesov2012,
  author  = {Kolesov, R. and Xia, K. and Reuter, R. and St{\"o}hr, R. and
             Zappe, A. and Meijer, J. and Hemmer, P. R. and Wrachtrup, J.},
  title   = {Optical detection of a single rare-earth ion in a crystal},
  journal = {Nature Communications},
  year    = {2012},
  volume  = {3},
  pages   = {1029},
  doi     = {10.1038/ncomms2034},
}

@incollection{gorller-walrand1998,
	title = {Chapter 167 {Spectral} intensities of f-f transitions},
	volume = {25},
	isbn = {978-0-444-82871-2},
	booktitle = {Handbook on the {Physics} and {Chemistry} of {Rare} {Earths}},
	publisher = {Elsevier},
	author = {Görller-Walrand, Christiane and Binnemans, Koen},
	month = jan,
	year = {1998},
	pages = {101--264},
}

@article{ozerov2019,
	title = {Computational study of the stable atomic trapping sites in {Ar} lattice},
	volume = {45},
	issn = {1063-777X},
	doi = {10.1063/1.5090045},
	number = {3},
	journal = {Low Temperature Physics},
	author = {Ozerov, Georgiy K. and Bezrukov, Dmitry S. and Buchachenko, Alexei A.},
	year = {2019},
	pages = {301--309},
}

@article{tam1993,
	title = {Matrix isolation spectroscopy of metal atoms generated by laser ablation. {III}. {The} {Na}/{Ar}, {Na}/{Kr}, and {Na}/{Xe} systems},
	volume = {99},
	issn = {0021-9606},
	doi = {10.1063/1.465348},
	number = {2},
	journal = {The Journal of Chemical Physics},
	author = {Tam, Simon and Fajardo, Mario E.},
	year = {1993},
	pages = {854--860},
}

@article{klotzbucher1997,
	title = {Holmium {Atoms} and {Small} {Clusters} in {Inert} {Matrices}},
	volume = {101},
	issn = {1089-5639},
	doi = {10.1021/jp970118a},
	number = {25},
	journal = {The Journal of Physical Chemistry A},
	author = {Klotzbücher, Werner E. and Petrukhina, Marina A. and Sergeev, Gleb B.},
	month = jun,
	year = {1997},
	pages = {4548--4554},
}

@article{kolachevsky2007,
	title = {Blue laser cooling transitions in {Tm} {I}},
	volume = {89},
	issn = {1432-0649},
	doi = {10.1007/s00340-007-2835-z},
	number = {4},
	journal = {Applied Physics B},
	author = {Kolachevsky, N. and Akimov, A. and Tolstikhina, I. and Chebakov, K. and Sokolov, A. and Rodionov, P. and Kanorski, S. and Sorokin, V.},
	month = dec,
	year = {2007},
	pages = {589--594},
}

@article{klotzbucher1994,
  author  = {Klotzb{\"u}cher, Werner E. and Petrukhina, Marina A. and Sergeev, Gleb B.},
  title   = {Reactions of Samarium Atoms in Inert and Reactive Matrices},
  journal = {Mendeleev Communications},
  year    = {1994},
  volume  = {4},
  number  = {1},
  pages   = {5--7},
  doi     = {10.1070/MC1994v004n01ABEH000323},
}

@article{battard2023,
	title = {Cesium atoms in a cryogenic argon matrix},
	volume = {108},
	doi = {10.1103/PhysRevA.108.042820},
	number = {4},
	journal = {Physical Review A},
	author = {Battard, T. and Lahs, S. and Crépin, C. and Comparat, D.},
	year = {2023},
	pages = {042820},
}

@article{lahs2026,
	title = {Fluorescence and relaxation dynamics of cesium in argon matrices: {Multiple} trapping sites and host-guest interactions},
	volume = {114},
	shorttitle = {Fluorescence and relaxation dynamics of cesium in argon matrices},
	url = {https://link.aps.org/doi/10.1103/4f6z-cm2t},
	doi = {10.1103/4f6z-cm2t},
	number = {3},
	urldate = {2026-11-23},
	journal = {Physical Review A},
	publisher = {American Physical Society},
	author = {Lahs, S. and Dinesan, H. and Mahapatra, S. and Chin, W. and Crépin, C. and Dontot, L. and Douady, J. and Gervais, B. and Comparat, D.},
	month = sep,
	year = {2026},
	pages = {032821},
}

@article{vutha_orientation-dependent_2018,
	title = {Orientation-dependent hyperfine structure of polar molecules in a rare-gas matrix: {A} scheme for measuring the electron electric dipole moment},
	volume = {98},
	shorttitle = {Orientation-dependent hyperfine structure of polar molecules in a rare-gas matrix},
	url = {https://link.aps.org/doi/10.1103/PhysRevA.98.032513},
	doi = {10.1103/PhysRevA.98.032513},
	number = {3},
	urldate = {2022-03-01},
	journal = {Physical Review A},
	author = {Vutha, A. C. and Horbatsch, M. and Hessels, E. A.},
	month = sep,
	year = {2018},
	pages = {032513},
}

\end{document}